\documentclass[12pt]{article}
\usepackage[centertags]{amsmath}
\usepackage{amsfonts}
\usepackage{amssymb}
\usepackage{amsthm}
\title{\bf Radiation reaction and renormalization
in \\classical electrodynamics of point particle\\   in any
dimension}

\author{P. O. Kazinski ${}^1$,
S. L. Lyakhovich ${}^2$
and A. A. Sharapov ${}^3$\protect\\
{\it Department of Physics, Tomsk State University,}\\
{\it Lenin Ave. 36, Tomsk, 634050 Russia}\\
e-mails: \\{${}^1$ kpo@phys.tsu.ru, ${}^2$ sll@phys.tsu.ru,
${}^3$ sharapov@phys.tsu.ru }}
\date{{\bf hep-th/0201046}}

\begin{document}
\def\thesection{\arabic{section}}
\def\theequation{\arabic{equation}}

\maketitle

\begin{abstract}
The effective equations of motion for a point charged particle taking
account of radiation reaction are considered in various space-time
dimensions. The divergencies steaming from the pointness of the particle are
studied and the effective renormalization procedure is proposed encompassing
uniformly the cases of all even dimensions. It is shown that in any
dimension the classical electrodynamics is a renormalizable theory if not
multiplicatively beyond $d=4$. For the cases of three and six dimensions the
covariant analogs of the Lorentz-Dirac equation are explicitly derived.
\end{abstract}


\section{ Introduction}

The problem of accounting for the radiation back-reaction to the
relativistic motion of a point charge has been the subject of intensive
studies since Dirac's seminal paper \cite{Dirac}. The equation of motion
proposed by Dirac coincides in nonrelativistic limit to the zero-size limit
of the Lorentz model for an electron \cite{Lorentz} and for this reason it
is often referred to as the Lorenz-Dirac (LD) equation. For modern review
and further references see \cite{LL,Jackson,TVW,Poisson,Edgren}. Apart from
the pure theoretical interest, the LD equation finds applications in physics
of accelerators and astrophysics \cite{LT}.

In this paper we formulate the general framework for deriving the LD
equation in arbitrary dimension space-time. The main problem one comes up
against when trying to consistently derive the effective equation of motion
for a point charge is the inevitable infinities arising due to the
``pointness'' of the particle. The elimination procedure for these
``classical'' divergencies is relied on the same renormalization philosophy
which is used in the quantum field theory and in this case one may put it to
a rigor mathematical framework. The point is that, as far as the classical
particle motion is concerned, one has to regularize the equations of motion
which are linear in the field. This may require to remove the infinities
only from a single Green function (that determines retarded Li\'enard and
Wiechert potentials), no products of such functions are needed. We show that
in this linear situation the regularization problem is resolved by the
standard tools of the classical functional analysis, without invoking to the
more powerful but less rigor machinery of the quantum renormalization
theory. As a result we establish the general structure of the Lorentz-Dirac
equations for any even $d$ as well as the counterterms needed to compensate
all the divergences. In the particular case of $d=6$ the results of our
analysis are in good agreement with those of work \cite{Kos} where
six-dimensional LD equation was obtained on basis of energy conservation and
reparametrization invariance arguments.

When this work had been completed, we learned about the paper
\cite{Gal'tsov} where the question was discussed of the radiation
reaction in various dimensions. In this paper, the distinctions are
noticed between even and odd dimensions and the radiation reaction
force is derived in $d=3$. It was also argued that d=3,4 are the only
dimensions where all the divergencies are removed by the mass
renormalization, that checks well with our analysis and previous
studies of the paper \cite{Kos}. It should be noted, however, that
the form of $3d$ integro-differential LD equation proposed in
\cite{Gal'tsov} differs in appearance from that derived in the
present paper.

The paper is organized as follows. In Section 2 we set notations and discuss
a difference between the retarded Green functions in odd and even
dimensions. The regularization procedure for the singular linear functionals
relevant to our problem is detailed in Section 3. In Section 4, this
technique is applied to deriving the LD equation in various dimensions.
Starting with case of even dimensions we develop a convenient regularization
scheme for deriving the four-dimensional LD equation and yet allowing to
generate its higher dimensional analogs by mere expansion in a
regularization parameter. The case of odd dimensions seems to be less
interesting as it leads to the {\it nonlocal} (integro-differential) LD
equation, so after the discussion of general structure of the LD force we
restrict ourselves by considering a simple example of $2+1$ particle. The
main result of this section is that in any dimension the infinities coming
from the particle's self-action can be compensated by a finite number of
counterterms added to the original action functional, that means the
renormalizability of classical electrodynamics. In concluding section we
summarize the results and outline the prospects of further investigations.

\section{ Equations of motion and Green function}

In this section we remind some basic formulas concerning the one-particle
problem of classical electrodynamics formulated in an arbitrary dimensional
space-time. The detailed treatment of the subject can be found, for example,
in \cite{IS}.

So, let ${\mathbb{R}}^{d-1,1}$ be $d$-dimensional Minkowski space with
coordinates $x^\mu $, $\mu =0,...,d-1$, and signature $(+-\cdots -)$.
Consider a scalar point particle of mass $m$ and charge $e$ coupled to
electromagnetic field. The dynamics of the whole system (field)+(particle)
is governed by the action functional
\begin{equation}
S=-\frac{N_d}4\int {d^dxF_{\mu \nu }F^{\mu \nu }}+e\int {d\tau A_\mu \dot
x\,^\mu }-m\int {d\tau \sqrt{\dot x^2}}\,\,\,,\;\;\;\;\;\;N_d=\frac{\pi ^{%
\frac{1-d}2}}2\Gamma \left( \frac{d-1}2\right) \,,  \label{act}
\end{equation}
where $F_{\mu \nu }=\partial _\mu A_\nu -\partial _\nu A_\mu $ is strength
tensor of electromagnetic field $A_\mu $ and dot means the derivative with
respect to the particle's proper time $\tau $\footnote{%
Strictly speaking, the parameter $m$ entering to the Lagrangian is, so
called, {\it bare mass}. The physical mass of the particle will be
introduced bellow within renormalization procedure.}. Hereafter we use the
natural system of units ($c=1,\hbar =1$) so that $[m]=-[x]=1$ and $%
[e]=1-[A]=(4-d)/2$.

The variation of the action ($\ref{act}$) results in a coupled system of the
Maxwell and Lorentz equations describing both the motion of a charged point
particle in response to electromagnetic field and the propagation of
electromagnetic field produced by the moving charge. In the Lorenz gauge $%
\partial ^\mu A_\mu =0$ the equations take form
\begin{equation}
\square A_\mu (x)=-N_d^{-1}j_\mu (x),\,\,\,\,\;\;\;\;D^2x^\mu ={\cal F}^\mu
\label{eq}
\end{equation}
where the right-hand sides of equations are given by the electric current of
the point particle moving along the world-line $x^\mu (\tau )$ and the
Lorentz force:
\begin{equation}
j^\mu (x)=e\int {d\tau \delta (x-x(\tau ))\dot x^\mu (\tau )}%
,\,\,\,\,\,\;\;\;\;{\cal F}_\mu =\frac emF_{\mu \nu }(x)Dx^\nu \,.
\label{jf}
\end{equation}
Hereafter we use an invariant derivative
\begin{equation}
D=\frac 1{\sqrt{\dot x^2}}\frac d{d\tau }  \label{der}
\end{equation}
whose repeat action $D^nx^\mu (\tau )$ on a trajectory remains intact under
reparametrization: $\tau \rightarrow \tau ^{\prime }(\tau )$.

It is well known that character of propagation of the electromagnetic waves
depends strongly on the space-time dimension $d$, and especially on its
parity. Mathematically, this manifests in quite different expressions for
the {\it retarded} Green function $G=\square ^{-1}$ associated to the
D'Alambert operator:
\begin{equation}
G(x)=\left\{
\begin{array}{ll}
\displaystyle \frac 12\pi ^{\frac{2-d}2}\theta (x_0)\delta ^{(d/2-2)}(x_0^2-%
{\bf x}^2) & {\rm for\quad d=4,6,8,\ldots } \\[5mm]
\displaystyle \frac{(-1)^{\frac{d-3}2}}2\pi ^{-\frac d2}\Gamma \left( \frac{%
d-2}2\right) \theta (x_0-|{\bf x}|)(x_0^2-{\bf x}^2)^{\frac{2-d}2} & {\rm %
for\quad d=3,5,7,\ldots }
\end{array}
\right. \,\,,  \label{gf}
\end{equation}
where ${\bf x}=(x^1,\ldots x^{d-1})$. In an even-dimensional space-time the
Green function is localized on a forward light-cone with the vertex at the
origin, while for the case of odd dimensions its support extends to the
interior of light-cone. These distinctions have a crucial physical
consequence which can be illustrated by the following gedanken experiment.
Suppose a source of light is turned on at an initial time $t$ and is turned
then off at a time $t^{\prime }$. If the number of space-time dimensions is
even, an observer, being located at some distance from the source, will see
the light signal with clear-cut forward and backward wave fronts separated
by the time interval $t-t^{\prime }$. In so doing, the magnitude of the
signal observed will not vary during the time provided the source works with
a constant intensity. This is in a good agreement with our daily experience.
Another picture would be observed in odd-dimensional space-time. Of course,
there will be a definite instant of time when the observer finds the source
to be turned on, it is the time point when the forward wave front reaches
his eyes, but thereafter he will see the source to go slowly out and no
sharply definite backward front will be observed. Thus, in odd-dimensional
Universe the light source being ones turned on can never be turned off!
Sometimes this phenomenon is mentioned as a failure of the Huygens principle
for odd-dimensions.

In the context of present work, the distinction just drawn will lead to
essentially different forms for the Lorentz-Dirac equation: it will be given
by a finite-order differential equation for even $d$'s and by an
integro-differential one for the odd ones.

Return now to the equations of motion in the one-particle problem. The
general solution is given to the field equation ($\ref{eq}$) by the sum of
its particular solution and the general solution of the homogeneous
equations
\begin{equation}  \label{heq}
\square A_{\mu}(x)=0\,,\;\;\;\;\;\;\;\;\;\;\; \partial^\mu A_\mu(x)=0\,,
\end{equation}
describing free electromagnetic waves incident on the particle. Using the
Green function ($\ref{gf}$) and the charge-current density vector ($\ref{jf}$%
) we can construct the particular solution as the retarded potentials of
Li\'enard and Wiechert,
\begin{equation}  \label{ret}
A_\mu(x)=-N^{-1}_d\int G(x-y)j_\mu(y)d^ny =-N_d^{-1} e\int G(x-x(\tau ))\dot{%
x}_\mu d\tau\,.
\end{equation}
Thus we arrive at unambiguously determined decomposition of the
electromagnetic potential into the exterior field ($\ref{heq}$) and the
field created by the particle ($\ref{ret}$). The combined action of these
fields on the particle is described by the Lorentz force ($\ref{jf}$), which
is composed of exterior electromagnetic field, if any, and an inevitable
action of the particle upon itself, i.e. Lorentz-Dirac force,
\begin{equation*}
{\cal F}_\mu={\cal {F}}^{ext}_\mu+{\cal F}^{LD}_\mu
\end{equation*}
The main technical and conceptual difficulty one faces with when calculating
${\cal F}^{LD}$ is the divergence of the integral for retarded field $(\ref
{ret})$ taken at the points of particle's trajectory - the fact is of no
great surprise if one bears in mind the singularity of the Coulomb potential
associated to a point charge. This problem is closely related to another
one, namely, the problem of electromagnetic mass of an electron and
sometimes the both are even identified.

In the next section we extend Dirac's result to an arbitrary dimensional
space-time renormalizing Green function. We will show that in higher
dimensions the elimination of infinities is not exhausted by renormalization
of mass parameter $m$ only but brings about new renormalization constants
having no analogs in the original theory ($\ref{act}$).

\section{ Regularization}

We start with some general remarks concerning the mathematical status of the
retarded Green function $G$. The Green function ($\ref{gf}$) being a kernel
of the inverse wave-operator (\ref{ret}) is well-defined when acting on a
smooth compactly supported source functions $j_\mu(x)$. The difficulties may
arise, however, in attempting to apply the operator to a singular current
like that produced by a point charge. The problem is twofold. First, except
for the case of $d=3$, the Green functions involve products of generalized
functions, namely, the derivatives of $\delta$-function multiplied by $%
\theta $-function. Such the products are ill-defined when one treats them to
be generalized functions of one variable, say $x^0$, considering the other
variables as parameters. It is the problem we will face with by restricting
the argument of Green function onto particle's world-line $x^{\mu}(\tau)$.
The second point concerns the geometry of the domain the Green function is
supported within. Depending of dimension, the support coincides with, or
bounded by, the light-cone surface
\begin{equation}  \label{lc}
x^2=0\,,\;\;\;\;\;\;\;x_0\geq 0\,,
\end{equation}
which is not differentiable at $x^\mu =0$. As a result, $G(x)$ is singular
at the vertex of light-cone and ill-defined even in a generalized sense.

Let us illustrate this point by a simple example which, however, will play a
significant role in subsequent analysis. Namely, consider the generalized
function $F(s)=\delta(s^2)$ defined on the real half-line $s\geq 0$
\footnote{%
When working with generalized functions one has to fix the dual space of
basic functions, and in fact the latter is an element of definition for the
former. The both are in a ``dialectical" relation to each other: extending
the space of basic function one narrows, at the same time, that of the
generalized ones and vice versa. The particular choice of a basic functional
space is mainly dictated by a problem to be considered. Hereafter one may
always thought of a basic function as any smooth function on real half-line.
In so doing, all the derivatives at zero are understood as the derivatives
on the right.}. Relations $s^2=0$, $s>0$ are the one-dimensional analog of ($%
\ref{lc}$). We put by definition
\begin{equation}  \label{def}
\delta[f]=\int_0^{\infty}\delta(s)f(s)ds=f(0)
\end{equation}
Then the integral associated to linear functional $F$ reads
\begin{equation}  \label{lf}
F[f]=\int_{0}^{\infty} \delta (s^2)f(s)ds=\frac{1}2\int_0^{\infty}
\delta(s)\left(\frac{f(\sqrt{s})} {\sqrt{s}}\right)ds=\frac{1}%
2\lim_{s\rightarrow +0}\frac{f(s)}{s}\,,
\end{equation}
where $f(x)$ is a test function. In general, this integral diverges, so that
functional $F$ is ill-defined. Note, however, that integral (\ref{lf}) does
have meaning when evaluating on basic functions vanishing at zero. The
question is how to extend the functional $F$, in a consistent way, from
subspace of functions vanishing at zero to the whole functional space. We
will call a solution to this problem as a regularization of generalized
function $F(s)$ and denote it by $\text{reg} F(s)$. For example, the
following expression solves the problem:
\begin{equation}  \label{reg}
\text{reg} F[f]=\int_{0}^{\infty } \delta(s^2)(f(s)-f(0))ds + a_0f(0)=\frac
12f^{\prime}(0)+a_0f(0)\,
\end{equation}
$a_0$ being arbitrary constant. Indeed, the functional ($\ref{reg}$) is
well-defined for any basic function $f(x)$ and coincides with ($\ref{lf}$)
if $f(0)=0$. In fact, ($\ref{reg}$) is the general solution to our problem
since the complementary space to that of vanishing-at-zero functions (i.e.
subspace on which $F$ comes to infinity) is one dimensional and spanned, for
example, by constant function $f(s)=1$. Regularizing linear functional $F$
we just replace the infinite value $F[1]=\infty$ by any finite constant $%
\text{reg}F[1]=a_0$.

This procedure can be straightforwardly extended to derivatives of $\delta$%
-function. Consider the generalized function
\begin{equation}  \label{dn}
F^n(s)=\delta^{(n)}(s^2)=\left(\frac{d}{ds^2}\right)^n
\delta(s^2)\,,\;\;\;\;\;\;\;\; n=1,2,\ldots\,.
\end{equation}
Then we put
\begin{equation}  \label{reg1}
\text{reg}F^n[f]=\int_{0}^{\infty}
\delta^{(n)}(s^2)\left(f(s)-f(0)-sf^{\prime}(0)-\ldots -\frac
{s^{2n}}{2n!}f^{(2n)}(0) \right)ds+
\end{equation}
\begin{equation*}
+a_0f(0)+a_1f^{\prime}(0)+\ldots +a_{2n}f^{(2n)}(0)=
\end{equation*}
\begin{equation*}
=\int_0^{\infty}\delta(s)\frac1{2s}\left(-\frac{d}{2sds}\right)^n\left(f(s)-%
\sum_{k=0}^{2n} \frac{s^k}{k!}f^{(k)}(0)\right)ds+%
\sum_{k=0}^{2n}a_{k}f^{(k)}(0)\,,
\end{equation*}
that yields
\begin{equation}  \label{reg2}
\text{reg}F^n[f]=\frac{f^{(2n+1)}(0)}{(n+1)!}+\sum_{k=0}^{2n}a_{k}f^{(k)}(0)%
\,.
\end{equation}
As before, the functional ($\ref{dn}$), as it stands, is defined only on
functions which are $o(s^{2n+1})$ as $s\rightarrow +0$. Subtracting from a
function $f$ several first terms of its Taylor expansion we just project $f$
onto the subspace of such functions. The value of functional on the
complementary ($2n+1$)-dimensional subspace is fixed by the arbitrary chosen
constants $a_0, \ldots, a_{2n}$.

The formula $(\ref{reg2})$ is a particular example of a quite general
mathematical procedure known as the regularization of divergent integrals or
generalized functions. The procedure is applicable to a wide class of
singular functionals, in particular, to functionals with polynomial
singularities \cite{GSh}.

For our purposes it is instructive to re-derive expression $(\ref{reg2})$ in
another, perhaps more familiar for physicists, way. Namely, using the
sequential approach to generalized functions one may represent $\delta$%
-function $(\ref{def})$ as
\begin{equation}  \label{seq}
\delta(s)=\lim_{a\rightarrow +0}\frac{e^{-\frac sa}}{a}\,.
\end{equation}
Substituting this representation to rel. (\ref{lf}) we get a one-parametric
family of well-defined functionals until $a\neq 0$:
\begin{equation}  \label{da}
F_a[f]=\int_0^{\infty}\frac{e^{-\frac{s^2}{a}}}{a}f(s)ds
\end{equation}
In so doing, $F[f]=\lim_{a\rightarrow 0}F_a[f]$. For this reason we refer to
$a$ as the regularization parameter. Given the basic function $f(s)$, the
integral (\ref{da}) defines a meromorphic function in $\sqrt{a}$ given by
the Laurent series
\begin{equation}  \label{ls}
F_a[f]=a^{-\frac12}\int_0^{\infty}e^{-t^2}f(t\sqrt{a})dt=\sum_{n=0}^{\infty}
a^{\frac{n-1}{2}}\frac{f^{(n)}(0)}{n!}\int_{0}^{\infty} e^{-t^2}t^ndt =
\end{equation}
\begin{equation*}
=\sum_{n=0}^{\infty}a^{\frac{n-1}{2}}\frac{f^{(n)}(0)}{n!}\Gamma\left(\frac{%
n+1}{2}\right)
\end{equation*}
In the limit $a\rightarrow 0$ of switching regularization off the
singularity of functional $F[f]$ appears as a simple pole, while the
non-vanishing term of the regular part of $(\ref{ls})$ coincides with (\ref
{reg}) if take $a_0=0$. Similarly, the delta-shaped sequence (\ref{seq})
defines a holomorphic function in $a$ with coefficients being the
generalized functions in $s$,
\begin{equation}  \label{dela}
\delta_a(s)=\frac{e^{-\frac sa}}{a}=\sum_{n=0}^{\infty}a^n\delta^{(n)}(s)\,.
\end{equation}
Substitution of the last expression to (\ref{da}) gives
\begin{equation}  \label{dna}
F^n[f]=\lim_{a\rightarrow 0}\frac1{n!}\frac{d^nF_a[f]}{da^n}
\end{equation}
We see that the regularization (\ref{reg2}) of functionals $F^n$ corresponds
to replacement
\begin{equation}  \label{rep}
a^{-\frac{k+1}{2}}\rightarrow a_k \,,\;\;\;\;\;\;\; k=0,1,\ldots, n\,,
\end{equation}
of all the poles in expression (\ref{dna}) by arbitrary finite constants.

Thus the sequential approach to the problem leads to the same finite part
for the functional to be regularized and contains the same ambiguity in the
final definition as the scheme based on the direct subtraction.

\section{ Renormalization and the Lorentz-Dirac force.}

Now we are going to explicitly calculate the Lorentz-Dirac force ${\cal F}%
^{LD}$. As the results drastically differ in odd and even dimensions,
we consider these cases separately.

\subsection{ Even-dimensional space-time.}

The retarded Green functions are given by the first line in (\ref{gf}). We
start with the case of $d=4$. The expression for LD force
follows from (\ref{ret}) and (\ref{jf}):
\begin{equation}
{\cal F}_\mu ^{LD}(s)=4e^2Dx^\nu (s)\int \theta (x^0(s)-x^0(\tau ))\delta
^{\prime }((x(s)-x(\tau ))^2)(x(s)-x(\tau ))_{[\nu }\dot x_{\mu ]}(\tau
)d\tau
\end{equation}
where square brackets mean antisymmetrization. Since for the massive
particle, equation $(x(s)-x(\tau ))^2=0$ means $x^\mu (s)=x^\mu (\tau )$
that in turn implies $s=\tau $ the integrand is supported at the point $%
s=\tau $. Changing the integration variable $\tau \rightarrow s-\tau $ we
get a singular integral of the form
\begin{equation}
{\cal F}_\mu ^{LD}(s)=4e^2Dx^\nu (s)\int_0^\infty \delta ^{\prime
}((x(s)-x(s-\tau ))^2)(x(s)-x(s-\tau ))_{[\nu }\dot x_{\mu ]}(s-\tau )d\tau
\end{equation}
and the singularity comes from the derivative of delta function. Following
to the general regularization prescription we replace $\delta ^{\prime }$ by
an appropriate sequence of smooth functions. For example, from (\ref{dela})
it follows that
\begin{equation}
\delta ^{\prime }(s)=\lim_{a\rightarrow +0}\frac \partial {\partial a}\frac{%
e^{-\frac sa}}a\,,\qquad s\geq 0\,.  \label{d'a}
\end{equation}
This leads to the regular expression for LD force
\begin{equation}
{\cal F}_\mu ^{LD}(s,a)=4e^2Dx^\nu (s)\frac \partial {\partial
a}\int_0^\infty e^{-\frac{(x(s)-x(s-t))^2}a}(x(s)-x(s-t))_{[\nu }\dot x_{\mu
]}(s-t)\frac{dt}a  \label{fa}
\end{equation}
\begin{equation*}
{\cal F}_\mu ^{LD}(s)=\lim_{a\rightarrow +0}{\cal F}_\mu ^{LD}(s,a)
\end{equation*}
Evaluating this integral by the Laplace method we get an asymptotic
expansion for ${\cal F}_\mu ^{LD}(s,a)$ in half-integer powers of $a$. The
actual calculations are considerably simplified if one notes that form of
the integral is invariant under reparametrizations. So, we may assume the
proper time $\tau $ to satisfy the additional normalization condition $\dot
x^\mu \dot x_\mu =1$. In this gauge, many of terms vanish in the Taylor
expansion for the exponential and pre-exponential factors in (\ref{fa}),
leaving us with
\begin{equation*}
(x(s)-x(s-\tau ))^2=\dot x^2\tau ^2-\frac 1{12}(D^2x)^2\tau ^4+\frac
1{12}D^2x\cdot D^3x\tau ^5+\left( \frac 1{45}(D^3x)^2+\frac 1{40}D^2x\cdot
D^4x\right) \tau ^6+O(\tau ^7)\,,
\end{equation*}
\begin{equation}
(x(s)-x(s-\tau ))_{[\nu }\dot x_{\mu ]}(s-\tau )=-\frac 12D^2x_{[\mu
}Dx_{\nu ]}\tau ^2+\frac 13D^3x_{[\mu }Dx_{\nu ]}\tau ^3-  \label{2}
\end{equation}

\begin{equation*}
-\left( \frac 18D^4x_{[\mu }Dx_{\nu ]}+\frac 1{12}D^3x_{[\mu }D^2x_{\nu
]}\right) \tau ^4+\left( \frac 1{30}D^5x_{[\mu }Dx_{\nu ]}+\frac
1{24}D^4x_{[\mu }D^2x_{\nu ]}\right) \tau ^5+O(\tau ^6)\,.
\end{equation*}
Substituting these expressions back to the eq.(\ref{fa}) we get an integral
of the form (\ref{ls}). The result of integration reads:
\begin{equation}
{\cal F}_\mu ^{LD}(a)=2e^2\dot x^2\sum_{k=2}^\infty (-1)^{k+1}\frac{k-1}{k!}%
\Gamma \left( \frac{k+1}2\right) f_\mu ^{(k)}(Dx,...,D^kx)a^{\frac{k-3}2}\,.
\label{ldf}
\end{equation}
The explicit expressions for the several first terms are given by
\begin{equation*}
f_{(2)}^\mu =D^2x^\mu ,
\end{equation*}
\begin{equation*}
f_{(3)}^\mu =D^3x^\mu -(D^2x)^2Dx^\mu
\end{equation*}
\begin{equation*}
f_{(4)}^\mu =D^4x^\mu -\frac 32(D^2x)^2D^2x^\mu -3D^2x\cdot D^3xDx^\mu
\end{equation*}
\begin{equation*}
f_{(5)}^\mu =D^5x^\mu -\frac 52(D^2x)^2f_{(3)}^\mu -\frac{15}2D^2x\cdot
D^3xD^2x^\mu -(4D^2x\cdot D^4x+3(D^3x)^2)Dx^\mu
\end{equation*}
\begin{equation*}
\cdots
\end{equation*}
\begin{equation*}
f_{(k)}^\mu =D^kx^\mu +\cdots
\end{equation*}
\begin{equation*}
\cdots
\end{equation*}
Notice that in view of reparametrization invariance of the model the vector
of the regularized LD force is transverse to the particle's velocity, i.e. $%
\dot x^\mu {\cal F}_\mu (s,a)=0$. Upon removing the regularization only the
first two terms of the series (\ref{ldf}) survive - one singular and one
finite - that leads to the well-known LD equation for a $d=4$ point charge
interacting with its own field
\begin{equation}
\left( m+\frac{e^2}2\sqrt{\frac \pi a}\right) D^2x^\mu =\frac 23e^2(D^3x^\mu
-(D^2x)^2Dx^\mu )\,,\qquad a\rightarrow 0\,.  \label{ld4}
\end{equation}
The r.h.s. of this equation describes a back-reaction of the particle upon
radiating of electromagnetic waves and this is more than just an
interpretation. Notice that our calculations, being as rigorous as possible,
contain, however, an apparent ambiguity related to the definition of Dirac's
$\delta $-function on half-line (\ref{def}). Indeed, starting with an
arbitrary delta-shaped sequence defined on the whole real line one may
restrict the respective functional to test functions vanishing identically
at $s<0$. The different sequences will then lead to the different results
distinguished from each other by an overall constant factor. For example,
any symmetric (with respect to zero) approximation for $\delta (s)$ will
give the additional $1/2$ multiplier in the r.h.s. of rel. (\ref{def}).
However, requiring the total energy of the system to conserve one has to
equate the work of the LD force to the energy of the electromagnetic field
{\it radiated} by accelerating particle that immediately leads to our
convention for $\delta $-function on half-line. It is the energy
conservation argument which is frequently used to derive the LD equation in
four dimensions (see e.g. \cite{Dirac}, \cite{Jackson}, \cite{Poisson}).

The physical interpretation for the infinite contribution in the l.h.s. of
eq. (\ref{ld4}) is also obvious: its appearance reflects the infinite
energy, or mass, of the field {\it adjunct} to the particle. Within the
paradigm of renormalization theory this singularity is removed by a simple
redefinition of the particle's mass: one just replace the sum of
unobservable bare mass $m$ and the infinite contribution due to the
interaction by a finite (experimental) value,
\begin{equation}
m_{exp}=m+\frac{e^2}{2}\sqrt{\frac{\pi}{a}}\,,
\end{equation}
so that renormalized action of $d=4$ particle takes the form
\begin{equation}  \label{d4}
S^{(4)}_{renorm}=\left( m_{exp}-\frac{e^2}{2}\sqrt{\frac{\pi}{a}}\right)\int
d\tau\sqrt{\dot{x}^2}
\end{equation}
This means that the classical electrodynamics in four dimensions is a
multiplicatively renormalizable theory.

The next task is to try to extend this result to the higher even dimensions
using the explicit expressions for the retarded Green functions (\ref{gf}).
To do this one has no need to repeat all the calculations from the very
beginning. In view of the relations (\ref{dela}) and (\ref{dna})
the desired
expressions for the LD force can be derived for any even $d$
by successive differentiation of the universal series (\ref{ldf})
obtained for $d=4$. More precisely,
\begin{equation}
{\cal {F}}_\mu^{(d)}=\frac{\sqrt{\pi }}{2\left( \frac{d-4}2\right) !}\Gamma
\left( \frac{d-1}2\right) \left( \frac \partial {\partial a}\right) ^{\frac{%
d-4}2}{\cal F}_\mu ^{LD}(a)|_{a=0}
\label{Fd}
\end{equation}
In the case of six dimensions the respective equation of motion reads
\begin{equation*}
\left( m-a^{-\frac 32}\frac{e^2\sqrt{\pi }}6\right) D^2x^\mu =\frac{4e^2}{45}%
\left( D^5x^\mu -\frac 53(D^2x)^2[D^3x^\mu -(D^2x)^2Dx^\mu ]-\right.
\end{equation*}
\begin{equation}
\left. -\frac{15}3D^2x\cdot D^3xD^2x^\mu -[4D^2x\cdot D^4x+3(D^3x)^2]Dx^\mu
\right) -  \label{ldf6}
\end{equation}
\begin{equation*}
-a^{-\frac 12}\frac{e^2\sqrt{\pi }}{16}\left( D^4x^\mu -\frac
32(D^2x)^2D^2x^\mu -3D^2x\cdot D^3xDx^\mu \right) \,.
\end{equation*}
Besides the infinite mass term we observe a new type divergence involving
fourth-order derivatives. It is interesting to note that both the
divergences are Lagrangian, i.e. they can be canceled out by adding
appropriate counterterms to the initial Lagrangian (\ref{act}), so that the
renormalized action reads
\begin{equation}
S_{renorm}^{(6)}=-\left( m_{exp}+a^{-\frac 32}\frac{e^2\sqrt{\pi }}6\right)
\int d\tau \sqrt{\dot x^2}-a^{-\frac 12}\frac{e^2\sqrt{\pi }}{32}\int d\tau
\sqrt{\dot x^2}(D^2x)^2  \label{ld6}
\end{equation}
Contrary to this, the finite part of the LD force containing fifth
derivatives of $x$'s cannot be represented as the variation of a
Lorentz-invariant functional. These results generally agree with the
previous analysis of work \cite{Kos}, where the explicit expressions for $6d$
LD force and the counterterm in (\ref{ld6}) were obtained from requirements
of energy conservation and reparametrization invariance \footnote{%
The radiation reaction force (\ref{ldf6}) differs from that of \cite{Kos} by
an overall coefficient. This seems because of minor inaccuracy in the
normalization of $6d$ Li\'enard-Wiechert potentials accepted in \cite{Kos}.}.

This situation is general and nothing changes this picture as the number of
dimensions increases. Namely, for any even $d$ the finite part of the LD
force is given by a polynomial function in the derivatives of $x^\mu (\tau )$
up to $(d-1)$ order inclusive. Note that the higher (odd) derivative $%
D^{d-1}x^\mu $ enters linearly to LD force and therefore no variation
principle for the LD equation exists. This reflects the dissipative
character of the system losing the energy due to the radiation. Besides,
there are $d/2-1$ divergent terms ''the most singular of which'' corresponds
to the infinite electromagnetic mass of the particle, while the
interpretation for the other terms is not so simple as the structures of
such types are lacking in the original theory. By analogy with the four and
six dimensions one may expect that all the singularities are Lagrangian and
can be removed by adding appropriate counterterms to the action (\ref{act}).
This appears to be the case. Indeed, since the Maxwell equations (\ref{eq})
are linear we may resolve them in a general form (\ref{ret}) and,
substituting result back to the action functional (\ref{act}), get a
functional of particle's trajectory only,
\begin{equation}
S=-m\int d\tau \sqrt{\dot x^2}-\frac 12N_d^{-1}\int d^dx\int d^dyj^\mu
(x)G(x-y)j_\mu (y)  \label{di}
\end{equation}
\begin{equation*}
=-m\int d\tau \sqrt{\dot x^2}-\frac{e^2}2N_d^{-1}\int ds\int d\tau \;\dot
x^\mu (s)G(x(s)-x(\tau ))\dot x_\mu (\tau )\,.
\end{equation*}
The first term is the usual action functional of free scalar particle and
the second describes self-action. Since the retarded Green function is
localized on the light-cone we can explicitly perform (after a suitable
regularization) one integration in the double integral and get the
Lagrangian model for the relativistic particle with higher derivatives. It is
not hard to check that in the cases of $d=4,6$ the regularized self-action
term in (\ref{di}) exactly reproduces the counterterms in the corresponding
action functionals (\ref{d4}), (\ref{ld6}). As to the LD force, it can not
be obtained from the higher derivative model since only symmetric part of
the retarded Green function actually enters to the nonlocal action (\ref{di}%
). It seems very likely that the same situation takes place in the higher
dimension as well.

To summarize, for even dimension $d>4$ one force to extend the original
Lagrangian of the free relativistic particle by the addition of $d/2-2$
extra (higher derivative) terms in order to get a renormalizable theory, so
that whole renormalization procedure involves, together with the physical
mass, $d/2-1$ arbitrary constants.

\subsection{ Odd-dimensional space-time}

The Green function is given by a product of $\theta -$ function and a
singular analytical expression (\ref{gf}). According to this, the gradient
of Green function entering to the strength tensor $F_{\mu \nu }$ of adjunct
electromagnetic field consists of structures proportional to $\theta $ - and
$\delta $ - functions. This allows us to decompose the respective LD force
onto the local and nonlocal parts and both of these parts are singular in
general. To remove the singularities one may apply the direct subtraction
scheme discussed above. The obtaining of explicit expressions requires,
however, a great amount of computations, so we confine ourselves by example
of $2+1$ particle to illustrate how this technique works in the most simple
case.

The local and nonlocal parts of the LD force are given by the integrals
\begin{equation}  \label{nl}
{\cal F}_{nonlocal}^\mu = e^2Dx_{\nu}(s) \int_{-\infty}^s d\tau \left(\frac
{(x(s)-x(\tau ))^{[\mu}\dot{x}(\tau )^{\nu]}}{|x(s)-x(\tau )|^3}\right)
\end{equation}
\begin{equation}  \label{l}
{\cal F}_{local}^\mu=-2e^2Dx_\nu (s)\int_{-\infty}^s d\tau \left(\frac{%
\delta ((x(s)-x(\tau ))^2)(x(s)-x(\tau ))^{[\mu}\dot{x}^{\nu ]}(\tau) }{%
|x(s)-x(\tau )|}\right)
\end{equation}
In near the coincidence limit $s\rightarrow \tau $ the numerator and
denominator of the integrand (\ref{nl}) behave like $(s-\tau)^2$ and $(s-\tau
)^3$, respectively, so that the integral diverges logarithmically, as one
would expect for the two-dimensional Coulomb potential. Using asymptotic
equality
\begin{equation}  \label{aeq}
2(x(s)-x(\tau ))^{[\mu}\dot{x}^{\nu ]}(\tau ) \sim D^2x^{[\nu}(s)Dx^{\mu
]}(s) (x(s)-x(\tau))^2\sqrt{\dot{x}^2(\tau )}\,,\,\,\,\, s\rightarrow \tau\,,
\end{equation}
we can extract the infinity as follows:
\begin{equation}  \label{extr}
{\cal F}^\mu_{nonlocal}=e^2Dx_{\nu}(s) \int_{-\infty}^s d\tau \left(\frac
{(x(s)-x(\tau ))^{[\mu}\dot{x}(\tau)^{\nu]}}{|x(s)-x(\tau)|^3}- \frac{%
D^2x^{[\nu}(s)Dx^{\mu ]}(s)\sqrt{\dot{x}^2(\tau )}} {2|x(s)-x(\tau)|}
\right) +
\end{equation}
\begin{equation*}
+ \delta m D^2x^\mu(s),
\end{equation*}
where
\begin{equation}
\delta m=\frac {e^2}2\int_{-\infty}^s \frac{d\tau \sqrt{\dot{x}^2(\tau )}}{%
|x(s)-x(\tau )|}=\infty\,.
\end{equation}
Now the first integral is regular and it describes the nonlocal self-action
of the point charge, whereas the second term gives rise to the infinite mass
renormalization. In fact, the form of counterterm is uniquely determined by
reasons of reparametrization invariance, right physical dimension and
short-distance behavior.

As to the local part of LD force (\ref{l}) it turns out to be finite and
proportional to the free equations of motion,
\begin{equation}
{\cal F}_{local}^\mu=\frac{e^2}2D^2x^\mu\,,
\end{equation}
Note that here we may not worry about the ``right" definition for the $%
\delta $-function on half-line as, in any case, this contribution is
absorbed by the renormalization of mass. It would be interesting to compare
the work produced by the renormalzed force (\ref{extr}) with the loss of
energy due to the radiation.

The main lesson to be learned from this consideration is that, despite the
nonlocal character of the self-action force in $2+1$ dimensions, the
divergent part of the LD force is local and even Lagrangian. There is no
doubt that the same conclusion remains true for any higher odd dimension.

\section{ Concluding remarks}

In this paper we have considered derivation of the Lorentz-Dirac
equation for a point charge in various dimensions.
In any even $d$
the radiation reaction is given by the general formula (\ref{Fd})
implying only formal differentiation of $4d$ LD force by
regularization parameter.  It has been shown that contrary to the
quantum electrodynamics in $d>4$, the classical electrodynamics of
point particles is a renormalizable theory, although
non-multiplicatively, beyond $d=4$. The necessity of extra
counterterms (involving higher derivatives) in addition to that
responsible for the mass renormalization (attributed naturally to the
infinite energy of field surrounding a point charge) seems rather
counterintuitive but this is a must for obtaining a reasonable
theory.

The results of the paper can be extended at least in two directions. First
one may consider a supersymmetric or higher-spin generalizations for the
relativistic particle. The models with $N$-extended world-line supersymmetry
have been studied in refs. \cite{GT,HPPT}. It has been shown that after
quantization these models can be consistently interpreted as relativistic
spinning particles of spin $N/2$. By analogy with the quantum field theory
one may expect that inclusion of supersymmetry will result in correction of
some singularities or will even lead to finite models. If this is the case,
the account of spin can give rise to a fully consistent theory of charged
point particles. The consistent interactions of massive arbitrary-spin
particles to exterior fields including higher dimensions have been
constructed in \cite{LShS,LShSh}. Although the spin induced radiation is
known, its back-reaction remains an open question even in $d=4$.

The second option is to extend the above analysis to p-brain system
universally coupled to a $(p+1)$-form field and other background fields. In this
set up one may rise a question of classical stability for such a system. At
the linear level the problem was studied in ref. \cite{VH}, where a full set
of constraints on masses and couplings was established for $0$-brain
minimally coupled to a multiplet of vector and scalar fields. It is
anticipated that generalizing this analysis to case of extended object we
get a certain restriction on background fields in a form of local equations of
motion. If so, this may shed new light upon the origin of low-energy
effective field equations in the string and brane world.

\section*{Acknowledgments}

We thank V. G. Bagrov and A. Yu. Morozov for useful discussions at Volga
Summer School, June 2001, where these results have been presented for the
first time. The work was partially supported by RFBR under the grant no
00-02-17-956, INTAS under the grant no 00-262 and Russian Ministry of
Education under the grant E-00-33-184. The work of AAS was supported by RFBR
grant for support of young scientists no 01-02-06420.

\end{document}